\documentclass[runningheads]{svjour2}

\smartqed  

\usepackage{graphicx}

\journalname{General Relativity and Gravitation}

\begin{document}

\title{Area spectrum and thermodynamics of KS black holes in Ho\v{r}ava gravity}

\author{Jishnu Suresh \and V C Kuriakose}

\institute{Jishnu Suresh \at
              Department of Physics \\
              Cochin University of Science and Technology\\
              Cochin - 682 022, Kerala, India\\
              \email{jishnusuresh@cusat.ac.in}
           \and
            V C Kuriakose \at
              Department of Physics \\
              Cochin University of Science and Technology\\
              Cochin - 682 022, Kerala, India\\
              \email{vck@cusat.ac.in}
}
\date{Received: date / Revised: date}

\maketitle

\begin{abstract}
We investigate the area spectrum of Kehagias-Sfetsos black hole in
Ho\v{r}ava-Lifshitz gravity via modified adiabatic invariant
$I=\oint p_i d q_i$ and Bohr-Sommerfeld quantization rule. We find
that the area spectrum is equally spaced with a spacing of $ \Delta
A=4 \pi l_p ^2$.  We have also studied the thermodynamic behavior of
KS black hole by deriving different thermodynamic quantities.


\end{abstract}

\section{Introduction}
Ho\v{r}ava proposed a renormalizable theory of gravity in four
dimensions. This is a non relativistic theory of gravity, which can
be considered as a candidate for General Relativity at UV scale
\cite{Horava1,Horava2,Horava3}. Various aspects and solutions of
this theory have been studied by many. By introducing a dynamical
parameter $\lambda$ in asymptotically Lifshitz spacetimes, a
spherically symmetric black hole solution was first given by $L\ddot
u$, Mei, and Pope \cite{LMP}. Cai, Cao, and Ohta studied the general
topological black holes in \cite{Caicaoohta}. In addition, Kehagias
and Sfetsos have found  a  black hole solution in asymptotically
flat spacetimes by choosing the dynamical parameter $\lambda=1$ and
introducing another  parameter $\omega$  \cite{KS}. Park obtained a
black  hole  solution  for arbitrary values  of the parameter
$\omega$ and the   cosmological  constant  $\Lambda_w$, with
$\lambda=1$  \cite{Park}. There are several studies on the
thermodynamic and cosmological properties of black holes in
Ho\v{r}ava-Lifshitz (HL) gravity theory
\cite{Park,Kofinas,Calcagni,Park1,Wei,nv1,nv2,Myung,Kim,Biswas,Eune}.
In this paper, we investigate the area spectrum of a black hole in
HL gravity.
\paragraph{}
It is widely believed that horizon area of a black hole is to be
quantized. In 1974 Bekenstein proposed that the  black hole area is
equally spaced \cite{Bekenstein1} and the discrete spectrum was
obtained as
\begin{equation}
 A_n=8\pi n l_p ^2 ,
\end{equation}
where $l_p$ is the Planck length and $n=1, 2, \dots$ . Bekenstein
also proved \cite{Bekenstein2} that the black hole horizon area
spectrum has a minimal spacing of $8\pi l_p ^2$. A method to
quantize the horizon area was put forward by Hod \cite{Hod1,Hod2} in
which the quasinormal modes (QNMs) were used. Considering both
quasinormal mode frequency and Bohr's correspondence principle he
found that the area spectrum is related to the real part of QNM.
Later Kunstatter \cite{Kunsatter,Vagenas} furthered this idea to
find, the action $I=\int dE/ \omega_R$ is an adiabatic invariant,
where $\omega_R$ is the real  part of QNM frequencies, which leads
to an equally spaced area spectrum. Maggiore provided a new
interpretation \cite{Maggiore,Vagenas1,Medved} in which he proved
that the physical frequency of QNM is determined by its real and
imaginary parts and he derived the area spectrum which is in
consistence with that of Bekenstein. Using these ideas, there are
several studies on the area spectrum of different kinds of black
holes \cite{Setare,Setare1,Setare2}.
 \paragraph{}
Recently Majhi and Vagenas \cite{Majhi} proposed another method to
quantize the horizon area without QNMs. In their work, adiabatic
invariant quantity connects to the Hamiltonian of the black hole and
using both Hamiltonian and Bohr-Sommerfeld quantization rule, they
derived the equally spaced entropy spectrum of a spherically
symmetric black hole. According to Jiang and Han \cite{Jiang}, the
adiabatic invariant quantity $\int p_i dq_i$, used in \cite{Majhi}
is not canonically invariant and they suggested that by using the
adiabatic invariant quantity of the covariant form  $I=\oint p_i d
q_i$, one can quantize the horizon area of a spherically symmetric
black hole.
 \paragraph{}
In this paper we adopt the methods suggested by Majhi and Vagenas
and Jiang and Han to quantize the horizon area. Therefore, by
introducing the adiabatic invariant and using Bohr-Sommerfeld
quantization rule, we determine the entropy spectrum and  the area
spectrum of a Kehagias-Sfetsos (KS) black hole in  HL gravity.
 \paragraph{}
The rest of this paper is organized as follows. In Sect. 2, we
investigate the thermodynamics of KS black hole in  HL gravity. The
phase transition and stability of the black hole are discussed. In
Sect. 3, the entropy and horizon area quantization via adiabatic
invariant and Bohr-Sommerfeld quantization rule are studied.
Finally, Sect. 4 ends up with a brief discussion and conclusion.

    \section{HL gravity and Thermodynamics}
Ho\v{r}ava has used the ADM formalism in which the four dimensional metric of general relativity is
parametrized as
\begin{equation} \label{admmetric}
 ds_4^2= - N^2  dt^2 + g_{ij} (dx^i - N^i dt) (dx^j - N^j dt),
\end{equation}
where $N$ and $N^i$ denote the lapse and shift functions, respectively. The Einstein-Hilbert action
is given by
\begin{equation}\label{einstein hilbert action}
\small S=\frac{1}{16 \pi G}\int d^{4}x \sqrt{g}~ N\left\{ \left(K_{i
j}K^{i j}-K^{2}\right)+R-2 \Lambda\right\},
\end{equation}
where $G$ is Newton's gravitational constant, $R$ is the curvature scalar and $K_{ij}$ is the extrinsic
curvature that takes the form,
\begin{equation} \label{extrinsic curvature tensor}
K_{ij}=\frac{1}{2 N}\left(\dot{g}_{i
j}-\nabla_{i}N_{j}-\nabla_{j}N_{i}\right),
\end{equation}
here the dot denotes  differentiation   with respect to the time
coordinate $t$. The action of non-relativistic renormalizable
gravitational theory proposed by Ho\v{r}ava can be written as
\cite{Horava1}
\begin{equation}\label{hl action}
I=\int dt~ d^{3}x \sqrt{g}~ N
\left(\mathcal{L}_{0}+~\mathcal{L}_{1}\right),
\end{equation}
 where
 \begin{equation}
  \mathcal{L}_{0}=\left\{\frac{2}{\kappa^{2}}\left(K_{i j}K^{i j}-\lambda K^{2}\right)+\frac{\kappa^{2}\mu^{2}
\left(\Lambda R-3
\Lambda^{2}\right)}{8\left(1-3\lambda\right)}\right\}
 \end{equation}
 and
 \begin{equation}
  \mathcal{L}_{1}=\left\{\frac{\kappa^{2}\mu^{2}
\left(1-4\lambda\right)}{32\left(1-3\lambda\right)}R^{2}-\frac{\kappa^{2}}{2\omega^{4}}Z_{i
j}Z^{i j}\right\}.
 \end{equation}
 Here,
 \begin{equation}
Z_{i j}=C_{i j}-\frac{\mu ~\omega^{2}}{2}R_{i j}~~,
\end{equation}
 in which $C_{i j}$ is the Cotton tensor and it has the form
 \begin{equation}\label{cotton tensor}
C^{i j}=\epsilon^{i k l}\nabla_{k}\left(R_{l}^{j}-\frac{1}{4}R ~
\delta_{l}^{j}\right) =\epsilon^{i k
l}\nabla_{k}R_{l}^{j}-\frac{1}{4}\epsilon^{i k j}\partial_{k}R ,
\end{equation}
where $\kappa^{2}$,~ $\mu$,~ $\omega$,~ $\lambda$ and $\Lambda$ are
constants. For spherically symmetric solution of HL gravity, let us consider the line element

\begin{equation} \label{lineelement}
ds^2 = -{N(r)}^2dt^2 +\frac{dr^2}{f(r)}+ r^2 d\Omega^2.
\end{equation}
Substituting this metric ansatz in Eq.(\ref{hl action}) and after
angular integration, the Lagrangian will reduce to
\begin{eqnarray}\label{modified lagrangian}
\tilde{{\cal L}}=\frac{\kappa^{2}\mu^{2}N}{8\left(1-3\lambda\right)\sqrt{f}}
\left\{\frac{\lambda-1}{2}f'^{2}+\frac{\left(2\lambda-1\right)
\left(f-1\right)^{2}}{r^{2}} \right. \nonumber \\
\left.
-\frac{2\lambda\left(f-1\right)}{r}f'-2\omega\left(1-f-rf'\right)\right\},
\end{eqnarray}
where
\begin{equation}
 \omega=\frac{8\mu^{2}\left(3\lambda-1\right)}{\kappa^{2}}.
\end{equation}

By giving $\lambda=1$ and solving the field equations obtained from
Eq.(\ref{modified lagrangian}), we can arrive at the
KS solution \cite{KS},
\begin{equation} \label{kssolution}
 f_{\textmd{\tiny{KS}}} =N^2_{\textmd{\tiny{KS}}}= 1 + \omega r^2-\sqrt{r(\omega^2 r^3 + 4\omega M)}.
 \end{equation}

\begin{figure}[h]
\centering
\includegraphics[width=0.65\columnwidth]{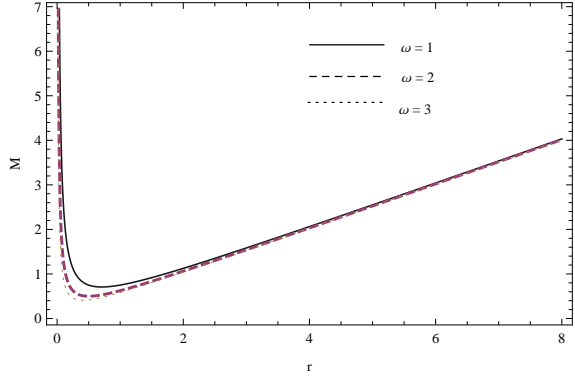}
\caption{Variation of mass with horizon radius for different values
of $\omega$.} \label{massentropy}.
\end{figure}

Now we will investigate the thermodynamic aspects of KS black hole.
In this section all quantities are expressed in the Planck
units($c=G=\hbar =1$). From the condition $f_{KS}(r_{\pm})=0$, the
outer and inner horizons are given by,
\begin{equation}\label{kshorizon}
r_{\pm}=M\pm \sqrt{M^2-\frac{1}{2\omega}}.
\end{equation}
By considering $r_{+}$ from (\ref{kshorizon}) we can establish a
connection between mass of the black hole and its horizon radius as,
 \begin{equation}\label{masshorizon}
  M=\frac{r_+}{2}+\frac{1}{4\omega r_+}.
 \end{equation}
From Bekenstein-Hawking area law, we can write
\begin{equation}\label{bhlaw}
 S=\frac{A}{4}= \pi r_+^2. 
\end{equation}
Hence, the horizon radius $r_+$ can be written in terms of entropy as,
\begin{equation}\label{rs}
 r_+=\sqrt{\frac{S}{\pi}}.
\end{equation}
Therefore, we can rewrite the mass-horizon radius (\ref{masshorizon}) as
\begin{equation}\label{msks}
 M=\frac{1}{4\omega}\sqrt{\frac{\pi}{S}}+\frac{1}{2}\sqrt{\frac{S}{\pi}}.
\end{equation}
As it is depicted in Fig.\ref{massentropy}, the two horizons of the
black hole merge at the point $r=r_e= 0.7$ (for $\omega=1$)
\cite{Myung1}. The same behavior is repeated for other values of
$\omega$ with slight changes in $r_e$ values. Thermodynamic
quantities such as temperature and specific heat are defined
respectively as,

\begin{equation}\label{tks}
 T=\left(\frac{\partial M}{\partial S} \right),
\end{equation}
\begin{equation}\label{cks}
 C=T\left(\frac{\partial S}{\partial T} \right).
\end{equation}

Then, from these equations, we can have the black hole temperature as
\begin{equation}\label{btks}
 T=\frac{1}{4\sqrt{\pi S}}-\frac{\sqrt{\pi}}{8\omega S^{\frac{3}{2}}}.
\end{equation}
and the heat capacity of the black hole as,
\begin{figure}[htbp]
  \begin{minipage}[b]{0.5\linewidth}
    \centering
    \includegraphics[width=\linewidth]{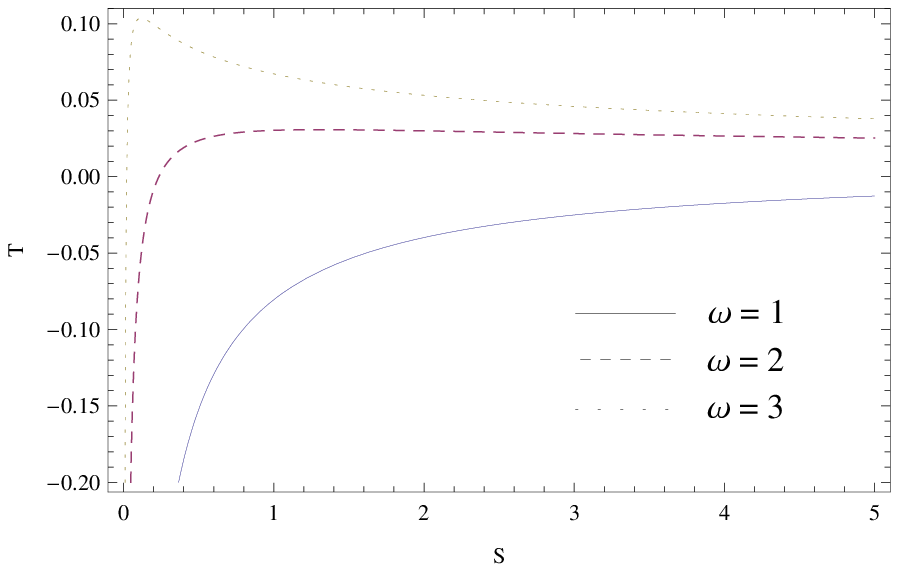}
    \caption{Variation of temperature with entropy for different values of $\omega$.}
    \label{tempentropy}
  \end{minipage}
  \hspace{0.5cm}
  \begin{minipage}[b]{0.5\linewidth}
    \centering
    \includegraphics[width=\linewidth]{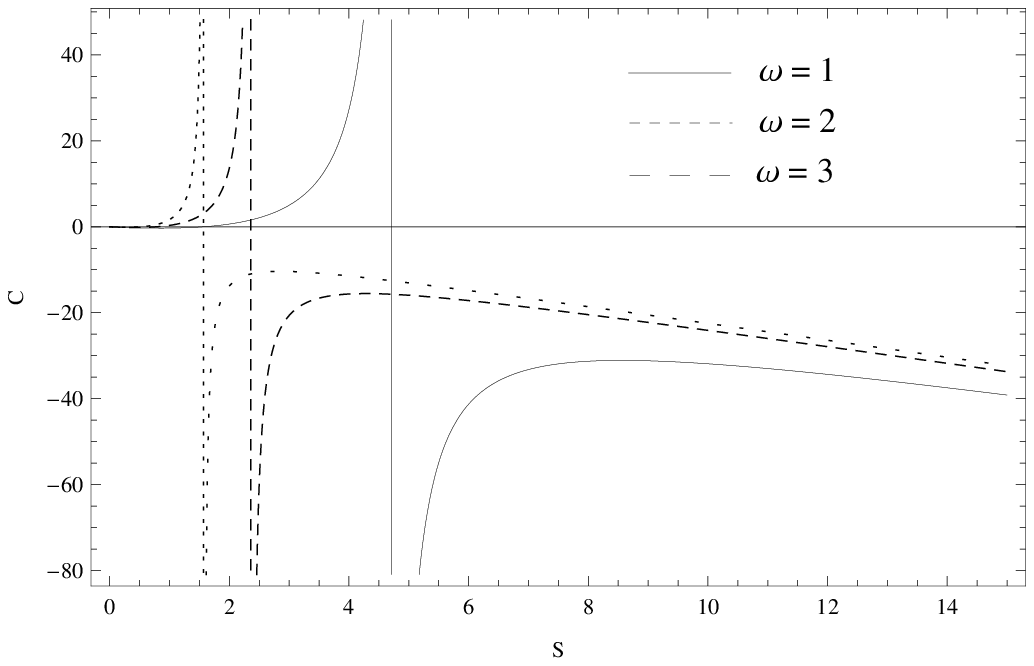}
    \caption{Variation of specific heat with entropy for different values of $\omega$.}
    \label{specentropy}
  \end{minipage}
\end{figure}

\begin{equation}\label{bhcks}
 C=- \left(\frac{4\omega S^2-2\pi S}{2\omega S-3\pi}\right).
\end{equation}
In Fig.\ref{tempentropy}, variation of temperature with respect to
entropy is plotted while in Fig.\ref{specentropy}, the variation of
heat capacity with respect to entropy for the different values of
coupling parameter $\omega$ is plotted. In Fig.\ref{specentropy},
there is a discontinuity in the plot, which shows that black hole
may undergo a  phase transition. Heat capacity is an important
thermodynamic quantity because from that we can tell about the
stability of the black hole. From Eq.(\ref{bhcks}) it is evident
that the heat capacity is positive for a range $\frac{\pi}{2\omega}
< S < \frac{3\pi}{2\omega}$. Hence, a KS black hole is stable for
this range  of  values  of S.

 \section{Area spectrum of KS black hole}
In this part of our work, entropy and horizon area of the KS black
hole are quantized via the adiabatic invariance and the
Bohr-Sommerfeld quantization rule. By considering the properties of
black hole such as adiabaticity and oscillating velocity of black
hole horizon, we can write the action as \cite{Jiang},
\begin{equation}\label{adiabatic covariant action}
 I=\oint p_i d q_i=\int_{q_{i} ^{in}} ^{q_{i} ^{out}} p_i ^{out} d q_i+\int_{q_{i} ^{out}} ^{q_{i} ^{in}} p_i ^{in} d q_i.
\end{equation}
Here $p_i ^{in}$ or $p_i ^{out}$ is the conjugate momentum
corresponding to the coordinate $q_i ^{in}$ or $q_{i} ^{out}$,
respectively, and $i= 0, 1, 2...$ It is also to be considered that
$q^{in}_{1}=r^{in}_{h}(q^{out}_{1}=r^{out}_{h})$ and $q_0 ^{in}
\left( q_0 ^{out} \right)=\tau$ where $r_h$ is the horizon radius
and $\tau$ is the Euclidean time with a periodicity
$\frac{2\pi}{\kappa}$ in which $\kappa$ is the surface gravity which
is given by,
\begin{equation}\label{kappa}
 \kappa=\frac{1}{2} \vline \frac{df_{\textmd{\tiny{KS}}}}{dr} \vline_{~ r_h}.
\end{equation}

Considering the Hamilton equation $\dot q_i = \frac{dH}{dp_i}$,
where $H$ is the Hamiltonian of the system, the integral given by
Eq.(\ref{adiabatic covariant action}),adiabatic covariant action can
be evaluated by considering the contour integration over a closed
path from $q_{i} ^{out}$ (outside the event horizon) to $q_{i}
^{in}$(inside the event horizon).

The action given by Eq.(\ref{adiabatic covariant action}) can be written as,
\begin{equation}\label{adiabatic covariant action parts}
 \int_{q_{i} ^{out}} ^{q_{i} ^{in}} p_i ^{in} d q_i= \int_{\tau_{out}} ^{\tau_{in}} \int_{0} ^{H} dH^\prime d\tau
 +\int_{r_{h} ^{out}} ^{r_{h} ^{in}} \int_{0} ^{H} \frac{dH^\prime}{\dot r_{h}} dr_h
  =2\int_{r_{h} ^{out}} ^{r_{h} ^{in}} \int_{0} ^{H} \frac{dH^\prime}{\dot r_{h}} dr_h~.
\end{equation}
where $r ^{out}$ and $r ^{in}$ denote the horizon location before
and after shrinking and $\dot r_h= \frac{dr_h}{d\tau}$ is the
oscillating velocity of black hole horizon. From the tunneling
picture, it is evident that when a particle tunnels in or out, the
black hole horizon will shrink or expand due to the loss or gain of
black hole mass \cite{Parikh}. Since tunneling and oscillation take
place at the same time we can write \cite{Zhang},
\begin{equation}\label{oscillating velocity}
 \dot r_{h}= -\dot r ~.
\end{equation}
where $\dot r$ is the velocity of tunneling particle.
Since the two contour integrals in the Eq.(\ref{adiabatic covariant action}) are equal we can write
it as,
\begin{equation}\label{adiabatic covariant action full}
\oint p_i d q_i=4\int_{r ^{out}} ^{r ^{in}} \int_{0} ^{H}
\frac{dH^\prime}{\dot r_{h}} dr_{h} ~.
\end{equation}

To evaluate this adiabatic invariant quantity for the black hole in
this discussion, let us consider the static spherically symmetric
space time given by the line element (\ref{lineelement}),

\begin{equation}
ds^2 = -{N(r)}^2dt^2 +\frac{dr^2}{f(r)}+ r^2 d\Omega^2~.
\end{equation}

To euclideanize this metric, we consider the transformation in time coordinate $t \rightarrow -i\tau$.
 Hence,
\begin{equation}\label{euclideanized lineelement}
 ds^2 = {N(r)}^2d\tau^2 +\frac{dr^2}{f(r)}+ r^2 d\Omega^2~.
\end{equation}

Let a photon travel across the black hole horizon, then from the
radial null path, i.e., $ds^2=d\Omega^2=0$ , we can write
\begin{equation}\label{radial null path}
 \dot r=\frac{dr}{d\tau}=\pm i \sqrt{N\left(r\right)^2 f\left(r\right)}~.
\end{equation}
From here onwards, our discussion will focus on the outgoing paths. From Eq.(\ref{kssolution}) it is evident that $f_{KS}=N_{KS}^{2}(r)$, and thus, from Eq.(\ref{radial null path}) we get
\begin{equation}\label{radial null path for ks}
 \dot r=+i f_{\textmd{\tiny{KS}}} \left(r\right)~.
\end{equation}
The adiabatic invariant will be,
\begin{equation}\label{adiabatic invariant integral for ks}
\oint p_i d q_i=-4i\int_{r ^{out}} ^{r ^{in}} \int_{0} ^{H} \frac{dH^\prime}{f_{\textmd{\tiny{KS}}} \left(r\right)} dr~.
\end{equation}
Using the near horizon approximation, $f_{KS} \left(r\right)$ can be Taylor expanded to get,
\begin{equation}\label{taylor series expansion}
 f_{\textmd{\tiny{KS}}} \left(r\right)=f_{\textmd{\tiny{KS}}} \left(r\right)~\vline~_{r_h}+(r-r_{h})\frac{d f_{\textmd{\tiny{KS}}} \left(r\right)}{dr}~\vline~_{r_h}+\cdots
\end{equation}
Since there is a pole at horizon $r_h$, we can consider a contour
integral over a half loop going above the pole from right to left.
Using the Cauchy's theorem, we can evaluate the integral in
Eq.(\ref{adiabatic invariant integral for ks}) to get
\begin{equation}\label{adiabatic invariant integral with kaapa and temperature}
 \oint p_i d q_i=4\pi \int_{0} ^{H} \frac{dH^\prime}{\kappa}=2 \hbar \int_{0} ^{H} \frac{dH^\prime}{T}~ ,
\end{equation}
where we have used the relation $T=\frac{\hbar \kappa}{2\pi}$ to connect temperature of the black hole
 with the surface gravity. According to first law of thermodynamics of black hole we can have,
\begin{equation}\label{first law of thermodynamics}
 dH^\prime=T dS~.
\end{equation}
Therefore,
\begin{equation}\label{adiabatic invariant integral final}
 \oint p_i d q_i=2 \hbar S~.
\end{equation}
From Bohr-Sommerfeld quantization
\begin{equation}\label{bohr sommerfeld quantization}
 \oint p_i d q_i=2\pi n \hbar, ~~~~~~~~ n=1,2,3,\cdots
\end{equation}
the entropy spectrum is given by
\begin{equation}\label{entropy spectrum}
 S=n \pi . ~~~~~~~~~~~~~~~~~ n=1,2,3,\cdots
\end{equation}
So the black hole entropy is quantized with a spacing of the entropy spectrum given by
\begin{equation}\label{entropy spectrum spacing}
 \Delta S=S_{\left(n+1\right)}-S_{\left(n\right)}=\pi.
\end{equation}
From Bekenstein-Hawking entropy relation \cite{Bekenstein2}, area spectrum can be found to be spaced as
\begin{equation}\label{area spectrum}
 \Delta A=4 \pi l_p ^2~.
\end{equation}
Thus,  we   see  that both  entropy and area spectra of KS black
hole  are quantized and  are equally spaced and they  are
independent of the black hole parameters.

\section{Discussion and conclusion}
In this paper we have studied the quantization of entropy and
horizon area of KS black hole in HL gravity using the method put
forward by Majhi and Vagenas, which was later modified  by Jiang and
Han. The entropy  and the area spectra are derived using  adiabatic
invariant and Bohr-Sommerfeld quantization rule and we have  showed
that both entropy and area spectra are equally spaced with a spacing
of $ \Delta S=\pi$ and $ \Delta A=4 \pi l_p ^2$ respectively.
Eventhough the  values of equispacing obtained  in the present study
are  different from the   values  obtained using  QNMs approach for
LMP black holes \cite{Majhi1} and KS black holes \cite{Momeni} in HL
theory, the equispaced property is  maintained and their order of
magnitudes are the same. Majhi \cite{Majhi1} has used tunneling
mechanism  and QNMs to study the entropy  spectrum of KS black hole
and found that though the  entropy  is quantized, the magnitude  of
equispacing obtained is different for  the  two  methods. The
results of the present study agrees with the result obtained through
tunneling mechanism  in  \cite{Majhi1}. The discrepancy in the
results may due to the fact that these methods are semiclassical. It is
also found that the  area  and  entropy  spectra  do not depend on
the black hole parameters.  We have also studied the thermodynamic
aspects of KS black hole and found that they are thermodynamically
stable \cite{Myung1} for a certain range of values of the entropy.

\begin{acknowledgements}.
The authors  are thankful  to  the  Reviewers for  valuable
suggestions. The authors wish to thank UGC, New Delhi for financial
support through a major research project sanctioned to VCK. VCK also
wishes to acknowledge Associateship of IUCAA, Pune, India.
\end{acknowledgements}

\end{document}